**Cybersecurity Information Sharing Governance Structures: An Ecosystem of Diversity, Trust, and Tradeoffs**

Elaine M. Sedenberg, UC Berkeley School of Information, esedenberg@berkeley.edu
James X. Dempsey, UC Berkeley School of Law, jdempsey@berkeley.edu

**1. Introduction**

Policymakers and corporate representatives have frequently discussed cybersecurity information sharing as if it were a panacea. The phrase itself refers to many different activities and types of exchanges, but from about 2009 to the end of 2015, the cybersecurity policy debate in Washington was dominated by calls for greater information sharing.[1] Influenced in part by the post-9/11 theme of "connecting the dots," both policymakers and the private sector commonly accepted that improved cybersecurity depended on—and would flow inexorably from—expanded information sharing within the private sector and between the private sector and the federal government.[2] This view seemed to rest upon the assumption that with more information, systems may be made more secure through prevention measures or rapid remediation. Policymakers, reluctant to regulate cybersecurity standards, viewed voluntary information sharing as a tangible coordination activity that could be incentivized through policy intervention and sometimes directly facilitated by federal government roles.[3] The policy debate culminated with the 2015 passage of the Cybersecurity Information Sharing Act (CISA).[4] The law sought to encourage information sharing by the private sector by alleviating concerns about liability for sharing otherwise legally restricted information. It also sought to improve sharing within the federal government and between the government and the private sector.

CISA was debated and adopted after several decades of efforts within law enforcement and national security agencies to coordinate and increase information sharing with and within the private sector. The US Secret Service (USSS) established the New York Electronic Crimes Task

Force (ECTF) in 1995 to facilitate information exchanges among the private sector, local and national law enforcement, and academic researchers. In 2001, the USA PATRIOT Act mandated that the USSS create a nationwide network of ECTFs, which eventually consisted of over 39 regional hubs.[5] In 1998, Presidential Decision Directive 63 (PDD-63) authorized the Federal Bureau of Investigation (FBI) to create a National Infrastructure Protection Center (NIPC) as a focal point for gathering and disseminating threat information both within the government and with the private sector.[6] PDD-63 simultaneously directed the national coordinator for infrastructure protection to encourage the private sector to create an Information Sharing and Analysis Center (ISAC).[7] The role of the private sector center was to collect and analyze private sector information to share with the government through the NIPC, but also to combine both private sector information and federal information and relay it back out to industry.[8] Although PDD-63 anticipated that there would be one national ISAC, various sectors ultimately formed their own ISACs focused on industry-specific security needs.[9]

      Over time, additional federal agencies also developed their own information sharing systems and procedures. For instance, US-CERT (US Computer Emergency Readiness Team)—an organization that took over many of NIPC's functions after it was dissolved following a transfer to the Department of Homeland Security (DHS)—releases vulnerability information and facilitates response to particular incidents. Various other information exchanges and feeds—each with its own scope, access policies, and rules—were established across federal agencies charged with securing aspects of cyberspace. For example, in 2001 the FBI formally announced its "InfraGard" project, designed to expand direct contacts with private sector infrastructure owners and operators, as well as to share information about cyber intrusions, exploited vulnerabilities, and infrastructure threats.[10]

In addition to these piecemeal federal efforts to expand cyber information sharing, private sector information sharing arrangements also proliferated. Antivirus software companies agreed to share virus signatures with each other, essentially deciding to differentiate themselves on platform usability and support instead of competing for data.[11] Additionally, security researchers and individual corporate professionals formed ad hoc arrangements around critical responses to major incidents such as the Conficker worm and the Zeus botnet—threats that required coordination of response as well as exchange of information.[12]

Consequently, even before CISA was enacted, an ecosystem of information exchanges, platforms, organizations, and ad hoc groups had arisen to respond to increasingly pervasive and complex security threats within all industries. Today, this ecosystem of information sharing networks is characterized by a high degree of diversity—the result of years of evolving policies and cooperative models, driven by both the federal government and private sector. Information sharing models and structures operate in different niches—working sometimes in silos, occasionally duplicating efforts, and sometimes complementing each other.[13]

CISA attempted to advance information sharing on four dimensions: within the private sector, within the federal government, from the private sector to the government, and from the government to the private sector. However, the legislation was enacted without first fully mapping the ecosystem that had developed in the preceding years. Little effort was made to identify what was working effectively and why, or to de-conflict existing federal programs. Instead, the private sector repeatedly stated—and policymakers accepted—that concerns over legal liability (mainly arising, it was asserted, from privacy laws) were inhibiting information sharing. Therefore, one of CISA's major provisions was liability protection for private sector organizations as an incentive for more information sharing.

CISA's usefulness and impact on the information sharing ecosystem has yet to be demonstrated. To the contrary, our study suggests that the law did little to improve the state of information sharing. If anything, it only added more hurdles to federal efforts by mandating that the federal portal include unnecessary technical details (free field text entry) and cumbersome submission methods (email). The law lacked specificity on how federal efforts would work with each other and with already existing information sharing networks in the private sector. Focusing almost solely on the private sector's liability concerns, it failed to address other key factors associated with sharing, including trust management, incentives, reciprocation, and quality control. In sum, CISA was a policy intervention divorced from existing sharing mechanisms and lacking a nuanced view of important factors that could enable agile exchanges of actionable information.

This paper focuses on cybersecurity information within the private sector and between the private sector and federal government (leaving to others the issue of sharing within the federal government itself). It examines how governance structures, roles, and associated policies within different cybersecurity information sharing organizations impact what information is shared (and with whom) and the usefulness of the information exchanged. This research is based on a qualitative analysis of 16 semi-structured interviews with cybersecurity practitioners and experts. Using these interviews and other available information on cybersecurity sharing, we have created a taxonomy of governance structures that maps the ecosystem of information sharing organizations—each of which fills particular security needs and is enabled by different policy structures. This paper discusses the implications of these policies and structures for values that directly impact sharing, particularly the tradeoff between trust and scalability. This research illustrates how different governance models may result in different degrees of success within the complex and changing cybersecurity ecosystem. Our findings point to lessons—mainly cautionary ones—for policymakers seeking to encourage improvements in cybersecurity. This paper focuses on information sharing within the United States, but given the multinational nature of many private sector companies, some findings may be relevant internationally.

The types of cybersecurity-related information that could be shared to improve cybersecurity defenses and incident response include incidents (including attack methods), best practices, tactical indicators, vulnerabilities, and defensive measures. Generally, the organizations we describe in this paper are engaged in sharing tactical indicators, often called "indictors of compromise" (IOCs). An IOC can be defined as an artifact that relates to a particular security incident or attack. IOCs may be filenames, hashes, IP addresses, hostnames, or a wide range of other information. Cybersecurity defenders may use IOCs forensically to identify the compromise or defensively to prevent it.[14]

**2. Taxonomy of Information Sharing Governance Structures and Policies**

Over time, different cybersecurity information sharing structures have arisen to address particular needs or challenges. Given the wide range of information types, federal roles, industry sectors, and information sensitivities at issue, it is perhaps inevitable that an array of information arrangements has formed, each serving particular perceived needs, each with its own priorities and challenges, and each with its own respective membership policies and governance structures. Our research identified at least seven information sharing models:

1) Government-centric
2) Government-prompted, industry-centric
3) Corporate-initiated, peer-based (organizational level)
4) Small, highly vetted, individual-based groups
5) Open-source sharing platforms
6) Proprietary products
7) Commercialized services

To understand these governance models, our taxonomy articulates different policy and organizational approaches to sharing, as well as their impact on mission, participation, risk/benefit tradeoffs, and efficacy.

Table 1: Taxonomy of Information Sharing Models

| Classification | Organizational Units | Example Organizations | Governance types |
|---|---|---|---|
| Government-centric | Government operated; private-sector members can be corporations, private sector associations (e.g., ISACs), nonprofits (e.g., universities), or individuals | DHS AIS; US-CERT; ECTF; FBI's e-guardian; ECS | Federal laws and policies; voluntary participation; rules range from open sharing subject to traffic light protocol or FOUO (for official use only) to classified information restrictions (ECS) |
| Government-prompted, industry-centric | Sector or problem specific | ISACs; ISAOs | Sector or problem specific; voluntary participation; generally organized as non-profits, use terms of service or other contractual methods to enforce limits on redisclosure of information |
| Corporate-initiated, peer-based (organizational level) | Specific private companies | Facebook ThreatExchange; Cyber Threat Alliance | Reciprocal sharing; closed membership; information controlled by contract (e.g., ThreatExchange Terms and Conditions) |
| Small, highly vetted, individual-based groups | Individuals join, take membership with them through different jobs | OpSec Trust; secretive, ad-hoc groups | Trust based upon personal relationships and vetting of members; membership and conduct rules |
| Open-source sharing platforms | | Spamhaus Project | Information published and open to all; no membership but may be formed around community of active contributors and information users; one organization may manage platform infrastructure |
| Proprietary products | Organization or individuals participate by purchasing the product | AV and firewall vendors | Information via paid interface; responsibility and security management still in house |
| Commercialized services | Organizations purchase service | Managed Security Service Providers | Outsourcing of security |

*2.1 Government-centric Sharing Models*

The cybersecurity policy of the US federal government is simultaneously oriented towards many different goals, ranging from national security, to protecting federal IT systems, to investigating and punishing cybercrime. with the overarching goal of ensuring a healthy and productive US economy through the protection of American critical infrastructures and intellectual property. Each goal results in different information sharing priorities.[15] Given the number of federal agencies involved in some aspect of cybersecurity, the growth of information sharing systems is not surprising—even if it is frustrating to information consumers. Federal information sharing programs range from the FBI's eGuardian and InfraGard, to DHS's Automated Information Sharing (AIS) program and its narrowly tailored Enhanced Cybersecurity Services (ECS) program, USSS ECTF alerts, and US-CERT alerts and tips.

The role of the federal government in improving cybersecurity may be viewed from a public good perspective, whereby federal investment in cybersecurity would adjust for underinvestment by individuals and the private sector.[16] However, for such public investment to be effective would require first an understanding of what the private sector lacks and whether the government has what is lacking or could effectively acquire it and make it available in a timely fashion. In fact, leaving aside the question of whether the private sector really suffers from a lack of cybersecurity information, there are limitations to the federal government's ability to quickly and efficiently share information. Accordingly, there are significant challenges associated with expecting the federal government to fulfill a role as central information collector and disseminator.

Given the network of national security, intelligence, and law enforcement entities, some government-held information becomes trapped within classification restrictions, involving extensive security standards for personnel, IT networks, and physical facilities, and severely limiting recipients and methods of disbursement. The Pentagon's Defense Industrial Base (DIB) cybersecurity program and DHS's Enhanced Cybersecurity Services (ECS) program were developed to disseminate such classified data within special security agreements. These programs trade limited access for greatly improved information quality. As implemented, they appear not to be intended to support dissemination to a wide number of recipients. Instead, they disseminate information to just a handful of communications service providers (AT&T, CenturyLink, and Verizon) plus Leidos (formerly SAIC), entities that provide cybersecurity services to a multitude of customers and have the capability to ingest and act upon the information provided. In contrast, the reach of DHS's Cyber Information Sharing and Collaboration Program (CISCP) is broader (although still restricted), but it focuses on sharing analytic products with the private sector and therefore trades speed for context. (CISCP offers a range of products, including indicator bulletins intended to support faster action to thwart attacks and remediate vulnerabilities.[17])

At the other end of the spectrum, membership requirements for organizations such as the USSS ECTFs are much less strenuous, requiring a referral by someone already in the organization. The ECTFs disseminate information mainly by email (and in-person meetings). Information shared on the listserv is regulated using the traffic light protocol, where each color defines how it may be used and re-disclosed.[18] Only the USSS sends information to the ECTF listservs, although the information may originate from many different sources.

Several interviewees discussed a hesitation after the Snowden revelations to share information with any US government agency, regardless of the formal governance mechanisms. They cited general cultural unease, as well as fear of negative publicity if and when the sharing came to light. One federal employee involved in information sharing commented that "post-Snowden, and almost certainly now post-WikiLeaks, [getting the private sector to share] is going to become more difficult for us. We are battling a lot of perception." Internationally, for any company subject to European regulation, these cultural and reputational concerns are heightened and augmented by the assumption that sharing information with the US government would violate the 2016 EU General Data Protection Regulation (GDPR).

The regulatory and law enforcement powers of the federal government at times may discourage sharing from the private sector. Yet the existence of those powers may also incentivize sharing, at least on a case-by-case basis, for they represent capabilities to act against cyberthreats in ways not available to the private sector.[19] One cybersecurity practitioner commented: "Law enforcement [are] the people who are able to take special action to identify and attribute this information to individuals, who have authority to utilize rule of law, court orders, subpoenas, everything that's required essentially to take authoritative action and prosecute these individuals. Nothing pulls them [attackers] out of the ecosystem quite as well as putting them in jail for their crimes."

Even when legal barriers and the government's negative reputation are mitigated, sharing with the government can be difficult. Interviewees complained that there is a high barrier to participation in DHS's AIS due to the technical requirements for setting up the sharing interface.

The fact that the US government's role in information sharing remains fractured among many different agencies—each with its own respective priorities to share inside or outside of the government itself—is not necessarily undesirable. It might be effective to have different agencies play different roles. However, it is not clear that there is a unified policy or strategy for the proliferation of federal information sharing programs with broadly defined and overlapping missions. What we see is a failure in both directions: private entities share relatively little information with the government, and what information the government shares is outdated or otherwise not actionable. Outside of specialized sharing arrangements such as the ECS, there are weak incentives for the private sector to take on the reputational risks and the administrative and technical burdens of sending information to the government.

Contrasted with the publicly endorsed but not yet realized goal of large-scale, large-volume sharing arrangements, the most effective reciprocal sharing between the private sector and the federal government may occur on an ad hoc basis, founded on personal connections between security professional in and out of government and on the unique strengths of particular agencies. For example, a national security agency may have the most to offer when an attacker is a foreign government, the FBI may have the most to offer when the attack appears to be a criminal matter, and the DHS or US CERT may be particularly useful in terms of remediation. In some reaches of the cybersecurity community, as one interviewee noted, there is a high crossover of personal relationships between "feds" and the private sector—which allows for direct sharing and consultation through interpersonal connections, as opposed to automated or systematic means. Given current trends, it seems there is a long way to go before the federal government could function as a central collector and switching hub for all cybersecurity information. Federal information sharing programs could benefit from a more realistic assessment of the federal government's strengths in partnering with, and responding to the needs of, the private sector.

*2.2 Government-Prompted, Industry-Centric Sharing Models*

As noted above, in 1998, President Bill Clinton directed his national coordinator for security and counter-terrorism to consult with owners and operators of critical infrastructure in order to encourage them to create "a private sector information sharing and analysis center." Although Clinton's directive contemplated a single center for all of the private sector, multiple ISACs were established over the next two decades, mainly on an industry-specific basis, to serve as collection and analysis points for private sector entities to share data on a peer-to-peer basis, to feed information into the federal government, and to provide a channel for federal information to flow out to the private sector. Though prompted by federal action, ISACs were intended to be led by the private sector. There are currently more than 20 ISACs. Their industry-specific focus seems to be based on the assumption that cybersecurity threats are most effectively shared among those within a single industry.[20]

In 2015, President Barack Obama encouraged the creation of Information Sharing and Analysis Organizations (ISAOs) to supplement the ISACs. This support for ISAOs was based in the belief that some companies do not fit neatly within a traditional industry classification.[21] ISAOs have sprung up around a variety of organizing principles, including industry (e.g., legal services, sports), region (e.g., Maryland, Southern California, Northeastern Ohio), or problem (e.g., Trustworthy Accountability Group (TAG), Cyber Resilience Group).[22]

The Financial Services ISAC (FS-ISAC) is widely cited as the canonical example of a successful information sharing arrangement. As of October 2017, it had 7,000 members, including commercial banks and credit unions of all sizes.[23] In its early days, the FS-ISAC benefitted from (among other factors) the financial sector's having a primary geographic hub, within New York City. Mutual dependencies among institutions in the financial services sector also helped supply the trust required to kick off the FS-ISAC. "The banks [although] competitors, are also counterparties. They know that even though they want to beat the other banks, they need them because they're on the other ends of the trades." Personal relationships between security professionals at the banks engendered trust. The importance of personal relations may have helped the FS-ISAC successfully navigate the hurdle of including law enforcement participants, by slowly introducing into the exchange "feds" who had existing relationships with members.

Trust based on geographic proximity and personal relationships has its limits. For the FS-ISAC, there may be a tradeoff between size and trust. It was reported in August 2016 that eight of the largest banks in the US had formed their own sub-group for cybersecurity information sharing and cooperation, one of "a couple dozen" sub-groups within or associated with the FS-ISAC.[24] Other factors associated with maturity may also impact trust. In 2016, the FS-ISAC sold its sharing platform, Soltra, to the for-profit NC4 because management had become too burdensome for the organization. Some of our interviewees expressed uncertainty about the seemingly sudden acquisition of the open-source platform by a proprietary company. Others viewed this as a sign of success, indicating that the FS-ISAC had matured to a point where its core platform could be commercialized.

From a governance perspective, ISACs and ISAOs represent a unique model: federal policy prompted their creation, but governance was ceded to voluntary groups of organizations facing common cybersecurity threats and sharing common goals. The hands-off, partnership model has fostered a network of organizations that responds to the needs and challenges of particular sectors, while offering the opportunity to coordinate with the federal government (e.g., many ISACs contribute to DHS's AIS and include law enforcement agencies within their membership). ISACs and ISAOs are typically nonprofits that manage membership and activities. For example, members of the FS-ISAC apply and pay a membership fee. Membership requirements vary by ISAC or ISAO, but the flexibility of the independent governance model allows each entity to reflect the needs of its community.[25] The National Council of ISACs (NCI) coordinates activities between ISACs and has a leadership presence at federal meetings, which helps to foster some high level collaboration.[26] An ISAO Standards Organization has also been set up, as a voluntary standard setting organization that works with information sharing entities on standards, guidelines, and best practices.[27] It is hard to assess the effectiveness of the ISACs and, even more so, the newer ISAOs. However, the federal government seems to have facilitated internal dynamics that allow trust to seed itself by encouraging the process of ISAC and ISAO creation but allowing industry to self-govern along sectoral or thematic lines.

*2.3 Corporate-Initiated, Peer-Based Groups*

Some companies have undertaken on their own initiative and without government intervention to coordinate information sharing in order to address particular needs. For example, antivirus vendors have agreed to share virus signatures and other indicators of compromise, essentially deciding to not compete on the underlying information but on other features of their products.[28] In 2014, these vendors formed the Cyber Threat Alliance (CTA). CTA requires that all participants contribute threat intelligence daily. It has designed a system that not only exchanges fresh indicators of compromise but also fosters discussions about the context for the shared data and produces "adversarial playbooks." Most recently, it has begun automating the delivery and configuration of endpoint controls on members' systems.[29] In 2017, CTA became a nonprofit and hired leadership to manage its growing network of participants—the same governance model by which many ISACs are run.[30]

Facebook's ThreatExchange also follows the closed membership and required participation model. ThreatExchange grew out of Facebook's efforts to rapidly handle malware spam attacks on its site that were also hitting other large internet companies.[31] Membership has been generally restricted to large peer companies, including Pinterest, Twitter, and Tumblr. Unlike ISACs, it is run by Facebook, not by an independent entity.

There is an unknown number of other privately-sponsored cybersecurity information sharing entities. The Advanced Cyber Security Center (ACSC), for example, was created by Mass Insight, a Boston-based consulting and research firm. ACSC brings together industry participants from the health care, energy, defense, financial services, and technology sectors, as well as government officials and academics. It is governed by a board of directors and a participation agreement, whereby members agree to share sensitive information confidentially.[32]

By orienting around a shared set of problems, these information exchanges can be tailored to fit the specific needs of their members (or their creators). As these exchanges appear to cater to larger, more established organizations, they may be better able to achieve reciprocity in sharing. However, this may leave out smaller companies, which need to find other means to secure their networks. In addition, these organizations may face issues of sustainability. For example, as of October 30, 2017, the most recent update to the homepage for Facebook's ThreatExchange was over a year old.

*2.4 Small, Highly Vetted, Individual-Based Groups*

Cybersecurity professionals have formed small, highly vetted online communities of peers to share sensitive, actionable information with the goal of promptly remediating attacks and other problems. Membership is held by individuals, not by organizations. These communities function largely in secret in order to protect their operations. Operations Security Trust is one example. Its skeletal website states: "Ops-T does not accept applications for membership. New candidates are nominated by their peers who are actively working with them on improving the operational robustness, integrity, and security of the Internet.[33] Though each group has its own membership vetting requirements and community standards, vetting usually involves a personal recommendation by a current member. Vetting rules may include restrictions on who can vet whom; some groups require that a newcomer have recommendations from individuals who do not work for the same employer as the newcomer. An interviewee noted: "Typically, the more vetted ones, they just don't typically vouch a lot of people that aren't under outside levels of trust already. With those, you may or may not get an invite. Then also, as far as kind of the vouching levels go, it may require zero vouches once you've been nominated ranging up to two or three different individuals who can vouch."

These groups are small by design, for their members require a high degree of trust in order to rapidly exchange information about ongoing attacks (which involves some disclosure of vulnerabilities), to solicit advice on how to respond, and to share lessons from attacks they have experienced (which again may involve some discussion of vulnerabilities) so others may take preventative actions. Interviewees stressed the importance of the small size of these organizations. For instance, one commented: "The unfortunate thing is the sliding scale, because as the groups become larger the pool of people may tend to start to evolve into a less trusting relationship because now there's more fingers in the pie, so to speak. You may not be quite aware of who your information is being disseminated to in some cases. More accidental or intentional or incidental leaks of information may occur as the constituency grows." The larger a group gets, the less likely it is to share sensitive information.

These clandestine and agile groups play an important role in the information ecosystem, allowing individuals to communicate quickly and completely with peers to actively mitigate incidents and devise preventative measures to protect their networks and systems. To the extent that members of these small, highly vetted groups participate in other sharing organizations with broader membership, they may help improve the functioning and effectiveness of those organizations (e.g., members of one of these small, highly vetted groups may share general knowledge with an ISAC or other sharing organization).

*2.5 Open Communities and Platforms*

Open-source sharing platforms and repositories for cybersecurity data offer a way to crowdsource collection, offer easy and unrestricted access to data, and allow for transparency and scrutiny of practices. Often associated with researchers (both independent and academic) or a host technology company, these platforms are most often focused on a particular type of data such as malware signatures or spam IP addresses. Policies about participation and use within these platforms and communities are generally liberal, and focused more on the structure and format of information shared.

Some of these open-source networks are run by nonprofits. The Malware Information Sharing Platform ("MISP") is a free, open-source platform for collecting, storing, and sharing cybersecurity indicators, initially developed by researchers from the Computer Incident Response Center of Luxembourg, the Belgian military, and NATO. Hail a TAXII.com is a repository of open-source cyberthreat intelligence feeds in STIX format, consisting of 817,631 indicators as of May 15, 2017.[34] The Spamhaus Project maintains the famous spam blocking list, which includes a Botnet Controller List, and also has an Exploits Block List, a real-time database of IP addresses of hijacked PCs infected by illegal third party exploits. Spamhaus disseminates intelligence on both a free and subscription basis.[35] There are also publicly available resources provided by for-profit entities for free, with various enhancements that users can purchase. For example, Snort, owned by Cisco Systems, is an open-source Network Intrusion Prevention System ("NIPS") and Network Intrusion Detection System ("NIDS") that performs real-time traffic analysis and packet-logging.[36] An exhaustive description of platforms and repositories that could fit within this category of open-sourced material is beyond the scope of this paper, but it is important to recognize how the openness and reach of communities like these differentiate them from more formal structures.

The governance of open-source initiatives has been widely studied, although it does not appear in the context of cybersecurity information sharing.[37] In 2012, DHS launched a project on open-source cybersecurity solutions, but the effort was not sustained and, in any case, did not address information sharing arrangements.[38] Hence, trust comes mainly from a belief in the value of transparency and the efficacy of the open-source model.

*2.6 Proprietary Products and Commercialized Services*

By "proprietary products," we refer to firewalls, antivirus software, and other software products that disseminate cybersecurity information through regular updates delivered to nodes or end user devices, often with little intervention by the network operator or end user. By "commercialized services," we refer to the wide range of outsourced cybersecurity services that use information they collect and analyze from existing information exchanges, from proprietary research, and from sensors embedded on customer networks to provide active monitoring and management of third-party devices and systems. Commercialized services include managed security service providers (MSSPs). Proprietary products and commercialized services represent models where information exchange has been commoditized by the market. Companies offering these products and services may participate in any of the other information exchanges, but they package and disseminate the information in a way that makes it available to small and medium organizations or individuals seeking to improve their security. Thus cybersecurity information from other sharing ecosystems reaches consumers or companies who may not have the security infrastructure in place to ingest and act on data feeds themselves. Additionally, these products and services may collect information from their customers and contribute these data back into the information sharing network, thus enlisting in the ecosystem entities that do not have the capability to collect or act upon information on their own.

From the end user perspective, proprietary products and commercialized services can be black boxes. Customers have no say in governance, and issues of trust are reduced to the single question of whether to purchase the product or service and to renew it when the initial contract term is up. It can be very hard for end users to make return on investment judgments, especially in the face of dynamic change in both the threat environment and the marketplace for these products and services. The Cyber Threat Alliance described above, now a nonprofit comprised of a dozen commercial entities, provides a form of governance, under which vendors commit to pool their intelligence. Ultimately, trade associations or other consortia may develop to offer other elements of governance.

**3. Discussion and Conclusions**

*3.1 Trust and the Tradeoffs*

The taxonomy of cybersecurity information sharing structures that we developed may help illustrate how different design and policy choices result in different information sharing outcomes. Based on the governance models described, we identified a set of factors or determinants of effectiveness that appear in different cybersecurity information sharing regimes.

The central role of trust in information sharing arrangements has been cited by many and is fully confirmed by our research.[39] Our research has identified one important aspect of trust: within cybersecurity information sharing, trust must be bidirectional. By this, we mean that 1) the sharing entity needs to trust that the information will not be used against it for regulatory or liability purposes, obtained by adversaries and exploited against it as a vulnerability, or disclosed publicly to hurt the reputation of the sharer; and 2) the recipient of information needs to trust the integrity of the information shared. We also found that success in some models has an additional dimension, which is reciprocity: parties need to trust that other participants will contribute roughly equivalent information. Governance structures and their associated policies may help generate these prerequisites by restricting and vetting membership in some capacity, by reviewing and verifying information submitted by other members, or by committing all members to contribute.

In the case of CISA, federal policy attempted to alleviate trust burdens that accompany sharing private sector information with the government, by limiting public disclosure through FOIA and by offering protections against liability and regulation. However, we found no evidence to indicate that CISA has succeeded in encouraging increased cybersecurity information sharing.[40] While it may be premature to conclude that CISA has been a failure, our research highlights some of the limitations of the statute's approach. By focusing on concerns over liability exposure, especially related to privacy laws, CISA failed to take into account other issues relevant to the sharing of private sector data with the federal government in a post-Snowden reality—particularly issues of public perception. Aside from the negative implications of sharing with the government, CISA did not account—and perhaps no law could account—for companies' fears about the reputational harm they might incur should their vulnerability become publicly known, or their fears about future attacks if vulnerabilities fall into the wrong hands. If indeed CISA has failed to induce more cybersecurity information sharing, it may be because it did not take into account these foundational elements of trust.

Our research points toward a clear tradeoff between membership size and the amount and sensitivity of information shared. Governance and policy structures can generate trust by limiting membership with some level of vetting and by requiring active participation. These dimensions of trust should be taken as governance design choices that can be worked into any organizational structure.

*3.2 The Ecosystem and the Role of the Federal Government*

The cybersecurity information sharing ecosystem, when considered in its entirety, shows the strengths of different components of the system. It is myopic to evaluate all the components of the system on the comprehensiveness of the information shared (let alone on timeliness or any other single metric). While it is necessary for at least some groups to have more complete or sensitive access to information, not every sharing organization in the ecosystem needs to have the same level of comprehensiveness or sensitivity. Each of the governing structures should be evaluated for success metrics that fit the goals of each model. For instance, by hosting regional, face-to-face meetings, ECTFs provide value in the ecosystem of information sharing and should not be pressured to be a primary distributor of all possibly relevant information. Ad hoc groups of highly vetted individuals, on the other hand, are not in competition with organizational-based systems. Nevertheless, overlap between individuals across types of groups can reinforce the ecosystem.

Proposals that the federal government should be the central collector and distributor of cybersecurity information seem unrealistic, if for no other reason than the trust deficit the government bears. Even if the government could satisfy the first two tenets of trust, a federally dominated exchange would only work if there were reciprocity between the federal government and the private sector. Current platforms like DHS AIS are struggling to distribute the information, not to mention the challenges brought on by the tradeoff of scalability and information comprehensiveness. Given classification concerns with most security-related data, it is unlikely that the federal government would ever achieve a fluid and agile reciprocation of information with the private sector.

Instead of suggesting the federal government as the central information hub for cybersecurity data, our research illustrates that other governing structures can fulfill necessary high-trust, high-sensitivity niches in the information exchanges. Certainly programs like CISCP and DIB allow the government to act in a way that fosters all the tenets of trust, but these programs will never be scalable without losing essential analytical resolution and reciprocity in sensitive sharing. Although still unsatisfying, the diverse economy of sharing models that we have identified may be, together and separately, the most feasible option.

---

[1] Cybersecurity information could refer to, for instance, malware signatures, suspicious IP addresses, or vulnerability disclosures. For statements supporting improved information sharing, see: Department of Homeland Security, "Information Sharing," accessed May 25, 2017, https://www.dhs.gov/topic/cybersecurity-information-sharing; Brad S. Karp, "Federal Guidance on the Cybersecurity Information Sharing Act of 2015," *Harvard Law School Forum on Corporate Governance and Financial Regulation*, 2016, https://corpgov.law.harvard.edu/2016/03/03/federal-guidance-on-the-cybersecurity-information-sharing-act-of-2015/; Evan McDermott and David Inserra, "Why Cybersecurity Information Sharing Is a Positive Step for Online Security," *The Daily Signal*, 2016, http://dailysignal.com/2016/01/25/why-cybersecurity-information-sharing-is-a-positive-step-for-online-security/.

[2] See, for example, Department of Homeland Security, "Cyberspace Policy Review: Assuring a Trusted and Resilient Information and Communications Infrastructure," 2009, i, https://www.dhs.gov/sites/default/files/publications/Cyberspace_Policy_Review_final_0.pdf.

[3] For instance, Senator Richard Burr (R-NC) said of CISA, "This landmark bill finally better secures Americans private information from foreign hackers…American businesses and government agencies face cyber-attacks on a daily basis. We cannot sit idle while foreign agents and criminal gangs continue to steal Americans' personal information as we saw in the Office of Personnel Management, Target, and Sony hacks." See Andrea Peterson, "Senate Passes Cybersecurity Information Sharing Bill Despite Privacy Fears," *The Washington Post*, October 27, 2015, https://www.washingtonpost.com/news/the-switch/wp/2015/10/27/senate-passes-controversial-cybersecurity-information-sharing-legislation/?utm_term=.d9b199957b9c.

[4] CISA, 129 Stat. 2936-2951, was enacted as part of the Cybersecurity Act of 2015, which itself was wrapped into the Consolidated Appropriations Act, 2016, Pub. L. 114-113 (December 18, 2015), https://www.congress.gov/114/plaws/publ113/PLAW-114publ113.pdf.

[5] US Secret Service. "The Investigation Mission," accessed May 25, 2017, https://www.secretservice.gov/investigation/; Department of Homeland Security, "United States Secret Service Electronic Crimes Taskforce," n.d.: 1–2, https://www.dhs.gov/sites/default/files/publications/USSS Electronic Crimes Task Force.pdf.

[6] "Presidential Decision Directive/NSC-63," 1998, https://fas.org/irp/offdocs/pdd/pdd-63.htm.

[7] Kristen Boon, Aziz Z. Huq, and Douglas Lovelace, eds., *U.S. Preparedness for Catastrophic Attacks* (New York, NY: Oxford University Press, 2012), 35, Google Books.

[8] It is interesting to note that the directive cites the Centers for Disease Control and Prevention (CDC) as a model for this type of information sharing.

[9] The ISACs focus on cybersecurity, but some have incorporated physical security into their missions.

[10] FBI National Press Office, "The FBI and the National Infrastructure Protection Center Publicly Introduce the National InfraGard Program," Washington, D.C., 2001, https://archives.fbi.gov/archives/news/pressrel/press-releases/the-fbi-and-the-national-infrastructure-protection-center-publically-introduce-the-national-infragard-program.

[11] Based on interviews and additional sources. For example, Cyber Threat Alliance, "A New Way to Share Threat Intelligence," accessed May 25, 2017, http://cyberconsortium.org/papers/Cyber_Threat_Alliance_White_Paper_9_5_2014.pdf.

[12] Andreas Schmidt, "Hierarchies in Networks: Emerging Hybrids of Networks and Hierarchies for Producing Internet Security," in *Cyberspace and International Relations: Theory, Prospects and Challenges* (Berlin, Heidelberg: Springer, 2013), 181–202, doi:10.1007/978-3-642-37481-4_11.

[13] Peter W. Singer and Allan Friedman, *Cybersecurity and Cyberwar: What Everyone Needs to Know*, Vol. 1, (Oxford: Oxford University Press, 2014), 224.

[14] Jason Andress, "Working with Indicators of Compromise," *ISSA Journal* (May 2015), https://c.ymcdn.com/sites/www.issa.org/resource/resmgr/journalpdfs/feature0515.pdf.

[15] Department of Homeland Security. "Cyberspace Policy Review: Assuring a Trusted and Resilient Information and Communications Infrastructure," 2009: 25, https://www.dhs.gov/sites/default/files/publications/Cyberspace_Policy_Review_final_0.pdf.

[16] Deirdre K. Mulligan and Fred B. Schneider, "Doctrine for Cybersecurity," *Daedalus* 140, no. 4 (2011): 1–30, doi:10.1162/DAED_a_00116.

[17] Department of Homeland Security, "Cyber Information Sharing and Collaboration Program (CISCP)," accessed May 25, 2017. https://www.dhs.gov/ciscp.

[18] United States Computer Emergency Readiness Team, "Traffic Light Protocol (TLP) Definitions and Usage," accessed May 25, 2017. https://www.us-cert.gov/tlp.

[19] CISA attempted to establish a shield for the private sector so that companies would not need to fear liability when reporting security information, but several interviewees remarked they did not feel the provision changed any perception about risks associated with sharing.

[20] The number of ISACs is based upon members of the National Council of ISACs. National Council of ISACs, "Member ISACs," accessed May 25, 2017, https://www.nationalisacs.org/member-isacs.

[21] Barack Obama, "Executive Order -- Promoting Private Sector Cybersecurity Information Sharing," *The White House Office of the Press Secretary*, 2015, https://obamawhitehouse.archives.gov/the-press-office/2015/02/13/executive-order-promoting-private-sector-cybersecurity-information-shari.

[22] ISAO Standards Organization, "INFORMATION SHARING GROUPS," accessed May 25, 2017, https://www.isao.org/information-sharing-groups/.

[23] See the FS-ISAC LinkedIn page (https://www.linkedin.com/company/fs-isac/) and "Testimony of John W. Carlson on behalf of the Financial Services Information Sharing & Analysis Center (FS-ISAC) before the U.S. House of Representatives Committee on Financial Services (June 24, 2015)," https://www.fsisac.com/sites/default/files/news/JCarlson%20June%2024%20Testimony%20FINAL.pdf.

[24] Robin Sidel, "Big Banks Team Up to Fight Cyber Crime," *Wall Street Journal*, August 9, 2016; Penny Crosman, "A Glimmer of Hope for Cyberthreat Data Sharing," *American Banker*, August 16, 2016, https://www.americanbanker.com/news/a-glimmer-of-hope-for-cyberthreat-data-sharing.

[25] FSIAC, "Membership Benefits," accessed May 25, 2017, https://www.fsisac.com/join.

[26] National Council of ISACs, "About NCI," accessed May 25, 2017, https://www.nationalisacs.org/about-nci.

[27] ISAO Standards Organization, "Information Sharing Groups About Us," accessed May 25, 2017 https://www.isao.org/about/.

[28] Eugene Kaspersky, "The Contemporary Antivirus Industry and Its Problems," *SecureList*, 2005, https://securelist.com/the-contemporary-antivirus-industry-and-its-problems/36063/.

[29] Rick Howard, "The Cyber Threat Alliance: How Far We've Come and Where We're Going," *Palo Alto Networks Blog*, 2017, https://researchcenter.paloaltonetworks.com/2017/02/cso-cyberthreat-alliance-far-weve-come-going/.

[30] Cyber Threat Alliance, "Cyber Threat Alliance Expands Mission through Appointment of President, Formal Incorporation as Not-for-Profit and New Founding Members," 2017, https://cyberthreatalliance.org/pr/pr-021317.html.

[31] Rex Santus, "Facebook's ThreatExchange Is a Social Platform for Sharing Cybersecurity Threats," *Mashable*, 2015, http://mashable.com/2015/02/11/threatexchange-facebook/#jnGMpTqOlZqb.

[32] Advanced Cyber Security Center (ACSC), "ACSC Membership," accessed July 1, 2017, https://www.acscenter.org/membership/.

[33] "Mission Operations Security Trust (or 'Ops-T')," accessed July 1, 2017, https://portal.ops-trust.net/.

[34] "Hail a TAXII," accessed July 1, 2017, http://hailataxii.com/.

[35] "The SpamHaus Project," 2017, https://www.spamhaus.org/.

[36] Cisco, "SNORT," *2017*, accessed July 1, 2017, https://www.snort.org; see also Gulshan Tweak, "Snort - What Is Snort (Network Intrusion Detection System)," *YouTube*, 2014. https://www.youtube.com/watch?v=S9J4SpbeJJE.

[37] See, for example, Siobhan O'Mahoney, "The governance of open source initiatives: what does it mean to be community managed?" *J. of Management and Governance* 11, no. 2 (June 2007), 139–150; Sonali K. Shah, "Motivation, Governance, and the Viability of Hybrid Forms in Open Source Software Development," *Management Science* 52, no. 7 (July 2006), 1000–1014.

[38] See DHS, Homeland Open Security Technology (HOST), https://www.dhs.gov/science-and-technology/csd-host.
23

[39] Matthew Harwood, "Lack of Trust Thwarts Cybersecurity Information Sharing," *Security Management*, 2011.

[40] None of our interviewees had observed a change in information sharing post-CISA. Several cybersecurity professionals we talked to said they were not aware of CISA; those who were aware of it indicated that it had not had any effect.

**Acknowledgements**

This work was funded by a grant from the Hewlett Foundation through the Center for Long-Term Cybersecurity (CLTC) at the University of California, Berkeley and was further supported by the National Science Foundation (NSF) Graduate Research Fellowship Program under Grant No. DGE1106400. Any opinions, findings, conclusions or recommendations expressed herein are those of the authors only and do not necessarily reflect the views of the Hewlett Foundation, the NSF, or the CLTC. We are grateful to the security experts and practitioners who spoke to us in the course of our research, and especially to Nicholas Weaver.